\documentclass[11pt]{article}

\usepackage{fullpage}

\usepackage{latexsym}
\usepackage{amssymb}   

\usepackage{amsmath}

\usepackage{enumerate}
\usepackage{mathtools}
\usepackage{verbatim}
\usepackage{color}

\def\e{\mbox{$\epsilon$}}

\def\X{\mbox{$\cal X$}}

\def\L{\mbox{$\cal L$}}

\def\F{\Bbb F}
\def\CN{\Bbb C}

\def\R{\Bbb R}

\def\cmp{\mbox{$\Bbb C$}}

\def\e{\mbox{$\epsilon$}}

\def\F{\Bbb F}
\def\Z{\Bbb Z}

\def\Z{\Bbb Z} 

\def\bind{B} 

\def\dmd{d'}

\def\finito{{\hspace*{\fill}  \mbox{$\blacksquare $}}}
\def\lemfinito{{\hspace*{\fill}  \mbox{$\blacktriangledown $}}}

\def\defeq{\stackrel{\mathrm{def}}{=}}

\def\ceil#1{\left\lceil #1 \right\rceil}

\newtheorem{theorem}{Theorem}[section]
\newtheorem{lemma}[theorem]{Lemma}
\newtheorem{definition}[theorem]{Definition}
\newtheorem{corollary}[theorem]{Corollary}
\newtheorem{conjecture}[theorem]{Conjecture}

\def\X{\chi}

\def\e{{{\epsilon}}}

\begin{document}
\author{Louay Bazzi
\footnote{
Department of Electrical and Computer Engineering, American University of Beirut, 
Beirut, Lebanon. E-mail: louay.bazzi@aub.edu.lb.} 
}
\title{Weight distribution of cosets of small codes  with good dual properties
\footnote{
  A preliminary version of this work  appeared in Proceedings of the 2015 IEEE International Symposium on Information Theory (ISIT 2015), Hong Kong, June 2015.}
\footnote{ 
  Two minor corrections which do not affect the validity of any of the the reported  results  were incorporated in July 2017.   
  Namely, Conjecture 3.4 in the older version is not correct.
  A counter example follows from Cohen's theorem (G. D. Cohen: A nonconstructive upper bound on covering radius, IEEE Trans. Inform. Th., vol. 29, no. 3, pp. 352-353, 1983)
  which asserts the
existence of  linear codes with covering radius  up to the sphere-covering bound.
The second correction is related to the ``Proof of Theorem 2.1 using Theorem 2.4'' in Section 2.  In that proof, 
the $n$-point Discrete Fourier Transform (DFT) should be  on $n+1$ points. 
The other steps of the proof hold without  modification.  
 }   
}

\maketitle


\begin{abstract}
The {\em bilateral minimum distance} of a binary linear code  is the maximum $d$ such that 
all nonzero codewords have weights between $d$ and $n-d$.  
Let $Q\subset \{0,1\}^n$ be a  binary linear code whose dual has bilateral minimum distance 
at least $d$, where $d$ is odd.
Roughly speaking, we  show that the average $L_\infty$-distance -- and consequently the $L_1$-distance -- between the 
weight distribution of a random cosets of $Q$ and the binomial distribution decays quickly as the bilateral minimum distance 
$d$ of the dual of $Q$ increases. For $d = \Theta(1)$, it decays like 
  $n^{-\Theta(d)}$. On the other $d=\Theta(n)$ extreme, it decays like 
and $e^{-\Theta(d)}$. 
It follows that, almost  all cosets of $Q$ have weight distributions  very close to the to the binomial distribution.
In particular, we establish the following bounds. 
If the dual of $Q$ has bilateral minimum distance at least $d=2t+1$, where $t\geq 1$ is an integer, 
then the average 
$L_\infty$-distance is at most $\min\{ 
\left(e\ln{\frac{n}{2t}}\right)^{t}\left(\frac{2t}{n}\right)^{\frac{t}{2} },
\sqrt{2} e^{-\frac{t}{10}}\}$. 
For the average 
$L_1$-distance, we conclude the bound $\min\{ 
(2t+1)\left(e\ln{\frac{n}{2t}}\right)^{t} \left(\frac{2t}{n}\right)^{\frac{t}{2}-1},
\sqrt{2}(n+1)e^{-\frac{t}{10}}\}$, 
which gives nontrivial results for $t\geq 3$. 
We given applications to the weight distribution of cosets of extended Hadamard codes and extended dual BCH codes.
Our argument is based on Fourier analysis, linear programming, and   polynomial approximation techniques. 
\end{abstract}


\section{Introduction}\label{introS}

The weight distribution of a random linear codes is well approximated by the binomial distribution 
$\bind_n(w) = \frac{\binom{n}{w}}{2^n}$ 
(see \cite{MS77}, page 287, and Lemma  \ref{avgl1} in this paper).
For nonrandom codes, the binomiality  of the weight distribution 
has been extensively studied in the  high rate regime, and applied to rate-$1$ BCH codes.   
 Strong approximation results were established 
in the literature  assuming that the dual code has good distance properties. 
Let $Q\subset \F_2^n$ be a block-length-$n$ $\F_2$-linear code and let 
 $Q^\bot$ be the dual of $Q$.  Let $\dmd$ be the minimum distance of $Q^\bot$,  and  let 
$\sigma$ be the {\em width} of $Q^\bot$, i.e., the minimum integer $\sigma$ such that 
$||y|-n/2|\leq \sigma/2$ for each nonzero $y\in Q^\bot$, where  $|y|$ denotes  the Hamming weight of $y$. 
Let $m_Q(0),m_Q(1),\ldots, m_Q(n)$ be the weight distribution of $Q$. That is,   
$m_Q(w)$ is the fraction of codewords of $Q$ of weight $w$ for $w=0,\ldots, n$
\footnote{In the literature, the weight distribution is usually used in reference  to  the {\em number} $A_Q(w)$  of codewords of $Q$ of weight $w$. Thus, in terms of $A_Q(w)$, $m_Q(w)=A_Q(w)/2^n$  is the {\em normalized} weight distribution.  For simplicity, we will refer to $m_Q$ as the weight distribution.}.  
The results in the literature can be divided into two 
categories: those assuming that the dual width $\sigma$ is small, and those assuming the weaker condition that the 
dual minimum distance $\dmd$ is large.

Assuming that the dual width $\sigma = o(n)$   and the rate $r$ of $Q$ is high (e.g., $r$ close to $1$), 
bounds of the form $m_Q(w) = \bind_n(w)(1+E_w)$ were established in 
\cite{S71,KFL85,S90,KL99}, where   
$E_n$ is the approximation error term.
This approach was initiated  by Sidel'nikov \cite{S71} who verified that the error term $|E_w| \leq n^{-0.1}$ if $r$ appropriately tends to 
 $1$.  
The bound on $|E_w|$ was later improved in \cite{KFL85,S90,KL99}, yielding exponential decay in $n$ for a certain range of $w$. 

In \cite{KL95,KL97,KL98,ABL01}, upper bounds on $m_Q(w)$ were established 
assuming that the dual distance $\dmd$ is large
 (linear in $n$) and the rate $r$ is high (the bounds at least require that $r$ does not go to zero as the block length $n$ 
increases).
Assuming that $\dmd=\alpha n$, for some constant $0<\alpha<1/2$, 
the bounds are of the form   
 $m_Q(w) = O(\bind_n(w)\sqrt{n})$ if 
$|w-n/2|\leq \beta$, for some constant $0 < \beta < 1/2$ which increases with $\alpha$. 
Unlike  upper bounds on  the dual width, lower bounds on the dual distance do not lead to 
lower bounds on the weight distribution (e.g.,
the code consisting of  even weight strings has dual distance $n-1$ but $m_Q(w) = 0$ for all odd $w$).
All the above bounds use MacWilliams's identity \cite{Mac63} and bounds on Krawtchouk polynomials. 

In \cite{Tie90,Tie91,S94,SS93}, lower bounds on the dual distance  have been also used  to derive upper bounds on the covering radius of the code, which 
is related to the width of the weight distribution. Another related work is \cite{GR08}, where worst case bounds on the moments of 
the weight distribution of cosets of dual BCH codes were derived 
based on the minimum distance of the dual code.

\subsection{Contribution}

In contrast with the above works on the binomiality of the weight distribution of codes, 
we  focus in this paper  on the low rate regime, and we study the weight distribution of a random  coset of the code   
rather than  of the code itself.   
Our bounds are for codes with small dual width $\sigma$, but  rather than formulating the statements in terms 
of dual width, we use the equivalent notion of dual bilateral minimum distance.
Define the {\em bilateral minimum distance}  of an $\F_2$-linear  code $C\subset \F_2^n$ to be the maximum $d$ 
such that  $d\leq |y|\leq n-d$, for each nonzero $y\in C$. 
We are interested in the bilateral minimum distance $d$ of the dual $Q^\bot$ of a given $\F_2$-linear code $Q \subset \F_2^n$.
Thus, $d$ is related to the width $\sigma$  of $Q^\bot$ via  $d=n/2-\sigma/2$. 
For technical convenience, we choose to express our results in terms of  $d$ rather than $\sigma$.  
We derive bounds which hold for values of $d$ 
as small as $d=3$   and as large as $d=\Theta(n)$.  
Note that if a  linear code $Q \subset \F_2^n$ is such that its dual   $Q^\bot$ has   bilateral minimum distance $d=\Theta(1)$, then 
the size of $Q$ is   typically $n^{\Theta(1)}$ (for random codes).  
On the other extreme, the 
$d=\Theta(n)$ regime typically corresponds to linear codes $Q \subset \F_2^n$ of size $2^{\Theta(n)}$. 
 
Roughly speaking, we  show that the average $L_\infty$-distance -- and consequently the $L_1$-distance -- between the   
weight distribution of a random cosets of $Q$ and the binomial distribution decays quickly as the dual bilateral minimum distance 
$d$ increases. For $d = \Theta(1)$, it decays like 
  $n^{-\Theta(d)}$.  On the other $d=\Theta(n)$ extreme, it decays like  $e^{-\Theta(d)}$. 
We given applications to the weight distribution of cosets of extended Hadamard codes and extended dual BCH codes, which fall in the  $d=\Theta(1)$ regime.

Before elaborating on the details of our bounds, it is appropriate to motivate them by comparing with the  above-mentioned  literature on the binomiality of the weight distribution of codes. 
As mentioned above, it is well known that the weight distribution of a random linear codes is well approximated by the binomial distribution. 
By a similar argument, if we fix any translation vector $u\in \F_2^n$ and choose a random linear code $Q \subset \F_2^n$ of a given dimension, 
the weight distribution of the coset $Q+u$  is well approximated by the binomial distribution (see Section \ref{discr}).
Our setup can be viewed a variant of this randomized scenario, where instead of fixing $u$ and choosing the code $Q$ at random, we fix the code 
and choose the coset (or equivalently $u\in \F_2^n$) at random.
 The advantage of our setup is that it requires much less randomness: we need 
$n-k$ random bits to choose a random coset of a given code  of dimension $k$,  
 compared to roughly $nk$ random bits needed to choose a random linear code of dimension $k$ 
\footnote{ 
More precisely, to choose a random binary linear code of dimension $k$, we need at least  $\log_2{\left((2^n-1)(2^n-2)\ldots (2^n-2^{k-1})\right)} - \log_2{\left((2^k-1)(2^k-2)\ldots (2^k-2^{k-1})\right)}$ random bits.
}. 
Our results say that if the code has large bilateral minimum distance, then the weight distribution of the random coset is  well approximated by the binomial distribution, 
i.e., almost all translations of the code have weight distributions similar to that of a random  code of the same size.
To compare our results with  the  above works on the binomiality of the weight distribution of non-random codes
\cite{S71,KFL85,S90,KL99,KL95,KL97,KL98,ABL01},  
 we note first that a key  common point between  our work and the previous works is the requirement on the dual code to have  good distance properties. 
The advantage of our results is that, unlike  the previous results, they are not restricted to rate-$1$ codes. Our bounds hold for codes of rates ranging 
from $0$ to $1$. In fact, as we explain below, they  are best seen in the rate-$0$ regime, and in particular in the context of codes of size $n^{\Theta(1)}$.
The advantage of the  previous results is that, unlike our results which require some randomness to choose the coset, they are applicable to the weight distribution of the
code itself.  
In summary, our setup can be seen as intermediate  scenario between the random-code scenario and the fixed-code  scenario: it is less random than the setup of  random codes, but 
it has enough randomness to establish strong bounds applicable in the the low rate regime.

We establish the following bounds.

 \begin{theorem}\label{sumth}
Let $Q\subsetneq \{0,1\}^n$ be an $\F_2$-linear code whose dual has bilateral minimum distance at least $d=2t+1$, where $t\geq 1$ is an integer. 
For each coset $Q+u$ of $Q$, where $u \in \F_2^n$, consider its weight distribution  
$m_{Q+u}(0),\ldots, m_{Q+u}(n)$. That is,  $m_{Q+u}(w)$ is the fraction of vectors in $Q+u$ of weight $w$ for $w=0,\ldots, n$.  
Let $U_n$ denote the uniform distribution on $\{0,1\}^n$ and let  ``$u\sim U_n$'' denote the process of 
choosing  and random vector $u$ according to $U_n$.

\noindent 
If $t\geq 1$, 
\begin{itemize}
\item[a)] {\em $($Small dual distance bound$)$} 
$E_{u\sim U_n} \| m_{Q+u} - \bind_n\|_\infty\leq \left(e\ln{\frac{n}{2t}}\right)^{t}\left(\frac{2t}{n}\right)^{\frac{t}{2} }.$

Thus, for $t = \Theta(1)$,  $E_{u\sim U_n} \| m_{Q+u} - \bind_n \|_\infty  =O( \frac{(\ln{n})^t}{n^{t/2}} )$. 
\item[b)] {\em $($Large dual distance bound$)$} 
$E_{u\sim U_n} \| m_{Q+u} - \bind_n\|_\infty\leq \sqrt{2} e^{-\frac{t}{10}}.$ 
\end{itemize}
If $t\geq 3$, 
\begin{itemize} 
\item[c)] {\em $($Small dual distance bound$)$} 
$E_{u\sim U_n} \| m_{Q+u} - \bind_n \|_1  \leq 
(2t+1)\left(e\ln{\frac{n}{2t}}\right)^{t} \left(\frac{2t}{n}\right)^{\frac{t}{2}-1}.$

Thus, for $t = \Theta(1)$,  $E_{u\sim U_n} \| m_{Q+u} - \bind_n \|_1  =O( \frac{(\ln{n})^t}{n^{t/2-1}} )$. 
\item[d)] {\em $($Large dual distance bound$)$} 
  $E_{u\sim U_n} \| m_{Q+u} - \bind_n \|_1  \leq  \sqrt{2}(n+1)e^{-\frac{t}{10}}.$
\end{itemize}  
For $n$ sufficiently large, the bounds in (a)  and (c) are better than those in  (b) and (d)  as long as 
 $\frac{d}{n} < \delta^*$,  where $\delta^*\approx 0.003446$.
\end{theorem}

%
%
%
%
%
%

%
%
%

The above bounds are best understood in the $d = \Theta(1)$ regime, which typically corresponds to codes of size
$n^{\Theta(1)}$.
The weight distribution of a random linear code $Q$  of size $n^{\Theta(1)}$ is 
$O(n^{-\Theta(1)})$-close to the binomial distributing (see Lemma \ref{avgl1}).
In the nonrandom case, 
the weight distribution of $Q$ may not be arbitrarily close to the binomial distribution even if the 
bilateral minimum distance of $Q$ is large. The simplest example is probably 
the extended Hadamard code $Q= H \cup (H + {\vec 1}) \subset \F_2^n$, where 
$H$ is the $(2^r-1,r,2^{r-1})$-Hadamard code, ${\vec 1}\in \F_2^n$ is the all-ones vector, $n=2^r-1$,  and $r\geq 2$ is an integer.
It is not hard to see that the size of $Q$ is $2(n+1)$ and that its  dual has bilateral minimum distance  greater than $d=3$ 
(see Section \ref{appS} for details). It has 
only $4$ possible weights $0,\frac{n-1}{2}, \frac{n+1}{2}$ and $n$,  thus 
$\|m_Q-\bind_n\|_\infty = \Theta(1)$
and $\|m_Q-\bind_n\|_1 = \Theta(1)$. 
However, by Part (a) of Theorem \ref{sumth} with $t=1$, we have  $E_{u\sim U_n} \| m_{Q+u} - \bind_n \|_\infty  =O( \frac{\ln{n}}{\sqrt{n}})$. 
It follows that for almost all cosets $Q+u$ of $Q$, we have $\| m_{Q+u} - \bind_n \|_\infty  =O( \frac{\ln{n}}{\sqrt{n}})$. 
Another example is the extended dual BCH code $Q$ 
of size $2(n+1)^t$, where $t\geq 3$ is a constant.   Its dual has 
bilateral minimum distance at least $2t+1$ but 
$\|m_Q-\bind_n\|_1 = \Theta(1)$ for all 
constant values of $t$ (see Section \ref{appS} for details).
However, by Part (c) of Theorem \ref{sumth}, we have 
$E_{u\sim U_n} \| m_{Q+u} - \bind_n \|_1  =O( \frac{(\ln{n})^t}{n^{t/2-1}} )$, hence 
for almost all cosets, we have $\| m_{Q+u} - \bind_n \|_1  =O( \frac{(\ln{n})^t}{n^{t/2-1}} )$.
 
\subsection{Proof technique}
   
At a high level, the proof of Theorem \ref{sumth} is based on Fourier analysis, linear programming, and   polynomial approximation techniques. 
Our argument is not based  on Krawtchouk polynomials, which  naturally arise when studying a property of the code  given dual constraints. 
In our problem,  we are studying an average over cosets  given dual constraints. This 
lead us to a  different type of approximations based on Taylor approximation of 
the exponential function.  
Unlike the above $L_\infty$ and $L_1$ statement in Theorem \ref{sumth}, our key technical result (Theorem \ref{main}) is   a mean-square-error statement.  
Using a squared norm  enabled us  to go the Fourier domain via  Parseval's equality.  
As in Delsarte's linear programming approach \cite{Del73},  
Linear programming naturally arise as a relaxation of the problem 
of optimizing over codes subject to  dual constraints 
to optimizing over probability distributions satisfying those constraints.  
Compared to the classical LP relaxations of coding problems, 
our  relaxation does not require the non-negativity of the Fourier transom of the 
probability distribution.  We use the code linearity before the relaxation via 
 Parseval's equality and a subtle application of  MacWilliams's identity.

\subsection{Original motivation}   
The original  motivation behind the work reported in this paper was the problem of explicitly constructing 
for each constant $c>0$  (a family of)  polynomial-size subset(s) $S\subset \F_2^n$ which is {\em pseudobinomial} in the sense that {\em for all} $u\in \F_2^n$, 
 the $L_1$-distance between the weight distribution of the translation of $S$ by $u$ 
is $n^{-c}$-close to the binomial distribution in the $L_1$-sense.  
 One consequence of the result in this paper is that polynomial-size 
linear codes with good dual properties achieve this goal for {\em almost all} $u\in \F_2^n$ 
(e.g., extended dual BCH codes).
We believe that the original goal which requires the stronger condition 
``{\em for all} $u\in \F_2^n$'' is not achievable using linear codes (see Section \ref{conjs}). 
In a recent paper \cite{Baz14},  we studied the notion of pseudobinomiality  in the context of small-bias probability distributions.
A probability distribution $\mu$ on $\{0,1\}^n$  is called 
 is {\em $\delta$-biased}
if  $|E_\mu \X_z|  \leq \delta$ for each nonzero $z\in \{0,1\}^n$,  where    $\X_x(x) \defeq (-1)^{\sum_i x_i z_i}$  \cite{NN93}. 
 Note that linear codes give rise to highly biased probability distributions because of their defining linear constraints. 
Namely, 
if $\mu_Q$ is the probability distribution resulting from choosing a uniformly random element of an $\F_2$-linear code $Q$, then   
 $E_{\mu_Q} \X_z=1$ for each $z\in Q^\bot$.  
If instead of a the linear code we have a $\delta$-biased probability distribution on $\{0,1\}^n$, using a
a much simpler argument, 
 bounds similar to those 
in Theorem \ref{sumth}  hold: $E_{u\sim U_n} \| \overline{\sigma_u \mu} -\bind_n\|_1 \leq \delta \sqrt{n+1}$, 
where $\overline{\sigma_u \mu}$ is the weight distribution of the $\F_2$-translation of $\mu$ by $u$, i.e., 
$\overline{\sigma_u \mu}(w) = \mu(x:|x+u|=w)$   (see Corollary 6.2 in \cite{Baz14}). 
The result in this paper can be seen as an extension of this bound to biased distributions. We elaborate on the comparison with 
small-bias probability distributions in Section \ref{sbias}.

\subsection{Paper outline}
In Section \ref{sresults}, we formally state our results and we reduce them to a mean-square error statement.  
We given in Section \ref{appS} applications to the weight distribution of cosets of extended 
Hadamard codes and extended dual BCH codes. In Section 
\ref{conjs}, we conjecture 
that there are no small codes such that the weight distribution 
of {\em all} cosets is arbitrarily close to the binomial distribution in the $L_1$-norm.
In Section \ref{frprelS}, we give some 
Fourier transform preliminaries used in the proof. 
The proof of our main technical result is in Section \ref{ProofS}. 
In Section \ref{sbias}, we compare with   small-bias probability distributions. 
In Section \ref{discr}, we compare our average $L_1$-approximation error with  random codes.

\section{Statement of the main results}\label{sresults}

If $x \in \{0,1\}^n$, the {\em Hamming weight} of $x$, which we  denote by $|x|$, 
 is the  number of nonzero  coordinates of $x$.     
If $C\neq 0\subset \F_2^n$  is an $\F_2$-linear code ($\F_2$ is the finite field  structure on $\{0,1\}$), 
the {\em minimum distance} of $C$ is the minimum weight of a nonzero codeword.  
Define the {\em bilateral minimum distance}  
of $C$ as the maximum $d$ 
such that  $d\leq |y|\leq n-d$, for each nonzero $y\in C$.
Note that, by definition, we must have $d\leq n/2$.  In most of our proposition, we will assume that 
$d\geq 3$, thus  $n\geq 6$.
A related notion is the 
{\em width} of a code $C$  (e.g., \cite{S90}), which is defined as   the minimum integer $\sigma$ such that 
$||x|-n/2|\leq \sigma/2$ for each nonzero $x\in C$. 
\footnote{In some works, e.g., \cite{KFL85}, the all-ones vector is allowed in $C$, i.e., 
the width of a  $C$   is  the minimum integer $\sigma$ such that 
$||x|-n/2|\leq \sigma/2$ for each  $x\in C$ other than zero and the all-ones vector.
We will not adopt this exception as our result rely on the all-ones vector not being present in the dual code.}
Thus, $d=n/2-\sigma/2$. For technical convenience, we choose to express our results in terms of 
bilateral minimum distance rather than width.

If $A\subset \{0,1\}^n$, let $\mu_A$ denote the probability distribution on $\{0,1\}^n$ uniformly distributed on $A$, i.e., 
\[
   \mu_A(x) \defeq\left\{\begin{array}{ll} \frac{1}{|A|} & \mbox{ if  } x\in A\\ 0 & \mbox{ otherwise.}\end{array}\right.
\]
We will denote the uniform distribution on $\{0,1\}^n$ by $U_n$, i.e., $U_n = \mu_{\{0,1\}^n}$. 
We will use the  notation  $[0:n] \defeq \{0, \ldots, n\}$.
If $\mu$ is a probability distribution on $\{0,1\}^n$, let $\overline{\mu}$ be the corresponding weight distribution on $[0:n]$, 
i.e., for all $w\in [0:n]$, 
\[
 \overline{\mu}(w)\defeq \mu(x \in \{0,1\}^n: |x|=w).
\] 
Thus, if $A \subset \{0,1\}^n$, then 
\[
\overline{\mu_A}(w) = \frac{|\{x\in A: |x|=w\}|}{|A|}
\]
is the fraction of elements of $A$ of weight $w$. 
 Note that $\overline{\mu_A}(w) = m_A(w)$  in terms of the  introductory notations used in Section \ref{introS}.
Let $\bind_n$ denote the binomial distribution on $[0:n]$, i.e., $\bind_n(w)=\frac{\binom{n}{w}}{2^n}$, and 
note that $\bind_n = \overline{U_n}$. 
Finally, if $\mu$ is a  probability distribution, 
let $E_\mu$  denote the expectation  with respect to $\mu$    and let ``$x\sim \mu$''  denote the process of sampling  a random vector $x$ according to $\mu$.

The following theorem is a restatement of Parts (a) and (b) of Theorem \ref{sumth}.

\begin{theorem}[$L_\infty$-bound]\label{cr1}
Let $Q\subsetneq \F_2^n$  be an $\F_2$-linear code whose dual $Q^\bot$ has bilateral minimum distance at least $d=2t+1$, where  $t\geq 1$  is an integer.
Then,   we have the bounds: 
\begin{itemize}
\item[a)] {\em $($Small dual distance bound$)$} 
\[
   E_{u\sim U_n} \| \overline{\mu_{Q+u}} - \bind_n\|_\infty
\leq \left(e\ln{\frac{n}{2t}}\right)^{t}\left(\frac{2t}{n}\right)^{\frac{t}{2} }. 
\]

\item[b)] {\em $($Large dual distance bound$)$} 
\[
   E_{u\sim U_n} \| \overline{\mu_{Q+u}} - \bind_n\|_\infty  \leq \sqrt{2} e^{-\frac{t}{10}}.
\]
\end{itemize}
\end{theorem}

An immediate consequence of the above  is the following corollary, which 
is restatement of  Parts (c) and (d) of Theorem \ref{sumth}.

\begin{corollary}[$L_1$-bound]\label{maincor}
Let $Q\subsetneq \F_2^n$  be an $\F_2$-linear code whose dual $Q^\bot$ has bilateral minimum distance at least $d=2t+1$, where $t\geq 3$  is an integer.
 Then, we have the bounds:  
\begin{itemize}
\item[a)]  {\em $($Small dual distance bound$)$} 
$$E_{u\sim U_n} \| \overline{\mu_{Q+u}} - \bind_n \|_1  \leq (2t+1)\left(e\ln{\frac{n}{2t}}\right)^{t}\left(\frac{2t}{n}\right)^{ \frac{t}{2}-1}.$$
\item[b)] {\em $($Large dual distance bound$)$} 
   $$E_{u\sim U_n} \| \overline{\mu_{Q+u}} - \bind_n \|_1  \leq \sqrt{2}(n+1)e^{-\frac{t}{10}}.$$ 
\end{itemize}
\end{corollary}
{\bf Proof:} The bounds  follow from Theorem \ref{cr1} since 
$\| \overline{\mu_{Q+u}} - \bind_n \|_1  = \sum_{w = 0}^n  | \overline{\mu_{Q+u}}(w) - \bind_n(w)| \leq (n+1) \| \overline{\mu_{Q+u}} - \bind_n\|_\infty.$ 
Note that in (a) we used the bound $n+1 \leq  d \frac{n}{d-1}$ (which holds for all $d\geq 2$ and $n \geq 1$ such that $d\leq n+1$), 
hence $(n+1)\left(\frac{2t}{n}\right)^{ \frac{t}{2}} \leq (2t+1) \left(\frac{2t}{n}\right)^{ \frac{t}{2} -1}$. 
Finally, note that the claim hold for all $t\geq 1$, but (a) gives  nontrivial bounds for 
$t\geq 3$.
\finito

\begin{corollary}\label{maincorc}
Let $Q\subsetneq \F_2^n$  be an $\F_2$-linear code whose dual $Q^\bot$ has bilateral minimum distance at least $d$, where $d\geq 7$ is odd. 
Assume that $d = \Theta(1)$.  Then for each $\e>0$,  $E_{u\sim U_n} \| \overline{\mu_{Q+u}} - \bind_n \|_1 \leq n^{- \frac{d-5-\e}{4}}$,  for $n$ large enough. 
Hence, for each $\xi>0$, for all but at most an  $n^{-\frac{\xi}{5}}$-fraction of the cosets $\{Q+u\}_u$, we have 
$\| \overline{\mu_{Q+u}} - \bind_n \|_1 \leq  n^{- \frac{d-5-\xi}{4}}$, for $n$ large enough. 
\end{corollary}
{\bf Proof:} This follows from Part (a) of Corollary \ref{maincor} and 
Markov Inequality.  
Namely, the probability, over the choice of $u\sim U_n$, that 
$\| \overline{\mu_{Q+u}} - \bind_n \|_1$ is larger than $n^{- \frac{d-5-\xi}{4}}$ is at most
\[
n^{\frac{d-5-\xi}{4}} E_{u\sim U_n} \| \overline{\mu_{Q+u}} - \bind_n \|_1  
\leq n^{\frac{d-5-\xi}{4}}d\left(e\ln{\frac{n}{d-1}}\right)^{\frac{d-1}{2}}
\left(\frac{d-1}{n}\right)^{ \frac{d-1}{4}-1} \leq n^{-\frac{\xi}{5}},
\]
for $n$ large enough.  
\finito

Our main technical result is Theorem \ref{main} below which unlike the previous statements is a mean-square-error statement. 

\begin{theorem}[Mean-square-error bound]\label{main}
Let $Q\subsetneq \F_2^n$  be an $\F_2$-linear code whose dual $Q^\bot$ has bilateral minimum distance at least $d=2t+1$, where $t\geq 1$ is an integer.
 If $0\leq \theta < 2\pi$, define $e_\theta:\{0,1\}^n \rightarrow {\Bbb C}$ by $e_\theta(x) = e^{i\theta |x|}$. Then,
for each $0\leq \theta < 2\pi$, we have the  bounds: 
\begin{itemize}
\item[a)]  {\em $($Small dual distance bound$)$} 
\[
   E_{u\sim U_n} | E_{\mu_{Q+u}} e_\theta - E_{U_n} e_{\theta}|^2 
\leq \left(e\ln{\frac{n}{2t}}\right)^{2t}\left(\frac{2t}{n}\right)^{t} 
\]
\item[b)]  {\em $($Large dual distance bound$)$} 
\[
   E_{u\sim U_n} | E_{\mu_{Q+u}} e_\theta - E_{U_n} e_{\theta}|^2 
  \leq 2 e^{-\frac{t}{5}}.
\]
\end{itemize}
\end{theorem}
The proof of Theorem \ref{main} is in Section \ref{ProofS}.   
We establish below  Theorem  \ref{cr1} using Theorem \ref{main}.

~\\  
\noindent
{\bf Proof of Theorem  \ref{cr1} using Theorem \ref{main}:} 
If  $w\in [0:n]$, define  the indicator function $I_w:\{0,1\}^n \rightarrow \{0,1\}$ by   $I_w(x) = 1$ iff $|x|=w$. 
Thus, $\bind_n(w)=E_{U_n}I_w$ and $\overline{\mu_{Q+u}}(w)=E_{\mu_{Q+u}} I_w$. 
For each $b \in [0:n]$, we have the character sum identity 
\[
\sum_{a =0}^{n} e^{\frac{2\pi i a b}{n+1}}  =\left\{\begin{array}{ll} n+1 & \mbox{ if  } b = 0\\ 0 & \mbox{ otherwise.}\end{array}\right.  
\]    
It follows that for each $x\in \{0,1\}^n$, 
\[
I_w(x) = \frac{1}{n+1} \sum_{a =0}^{n} e^{\frac{2\pi i a (|x|-w)}{n+1}}  =\sum_{a =0}^{n} \alpha_{a,w}  e_{\theta_a}(x),    
\]    
where $\alpha_{a,w} = \frac{1}{n+1}e^{\frac{-2\pi i wa}{n+1}}$ 
and $\theta_a = \frac{2\pi  a}{n+1}$. 
 Thus, for all $w\in[0:n]$ and $u\in \{0,1\}^n$, we have
\begin{eqnarray*}
|\overline{\mu_{Q+u}}(w)- \bind_n(w)| &=& | E_{\mu_{Q+u}} I_w - E_{U_n} I_w| 
= |\sum_{a=0}^{n} \alpha_{a,w}  (E_{\mu_{Q+u}} e_{\theta_a} - E_{U_n} e_{\theta_a})|\\
 &\leq&  \sum_{a=0}^{n} |\alpha_{a,w}| |E_{\mu_{Q+u}} e_{\theta_a} - E_{U_n} e_{\theta_a} |=
 \frac{1}{n+1}\sum_{a=0}^{n}  |E_{\mu_{Q+u}} e_{\theta_a} - E_{U_n} e_{\theta_a} |.
\end{eqnarray*}
By Jensen's inequality,   
$\left( E_{u\sim U_n} |E_{\mu_{Q+u}} e_{\theta} - E_{U_n} e_{\theta}|\right)^2 \leq  E_{u\sim U_n} |E_{\mu_{Q+u}} e_{\theta} - E_{U_n} e_{\theta}|^2$,
or any $0\leq \theta <2\pi$. It follows that:
\begin{eqnarray*}
E_{u\sim U_n} \|\overline{\mu_{Q+u}}- \bind_n\|_\infty & = &  E_{u\sim U_n} \max_{w}|\overline{\mu_{Q+u}}(w)- \bind_n(w)|\\
                                                & \leq & E_{u\sim U_n}  \frac{1}{n+1}\sum_{a=0}^{n} |E_{\mu_{Q+u}} e_{\theta_a} - E_{U_n} e_{\theta_a} |\\
                                                & =&                                                \frac{1}{n+1}\sum_{a=0}^{n} E_{u\sim U_n}   |E_{\mu_{Q+u}} e_{\theta_a} - E_{U_n} e_{\theta_a} |\\                        
                                                &\leq &   \max_{\theta}  |E_{\mu_{Q+u}} e_{\theta} - E_{U_n} e_{\theta} |\\                        
                                                &\leq &   \max_{\theta} \sqrt{E_{u\sim U_n} |E_{\mu_{Q+u}} e_{\theta} - E_{U_n} e_{\theta}|^2}. 
\end{eqnarray*}
Theorem 
  \ref{cr1} then follows from Theorem \ref{main}.
\finito

\section{Applications}\label{appS}
In this section, we apply Theorem \ref{cr1} to  the weight distribution of cosets of extended 
Hadamard codes, and  Corollary \ref{maincor}  to  the weight distribution of cosets of extended  dual BCH code. 
We conclude with a   conjecture that
 there are no small codes such that the weight distribution 
of {\em all} cosets is arbitrarily close to the binomial in the $L_1$-norm.

A natural construction of codes with large dual bilateral minimum distance  from codes with large minimum distance is the following.
\begin{lemma}\label{biconst}
If $n$ is odd and $D \subset \F_2^n$ is an $\F_2$-linear code 
of minimum distance at least $d$  such that the all-ones vector ${\vec 1} \in D$. 
Then  $Q\defeq D^\bot \cup (D^\bot + {\vec 1})$  is a 
code whose dual has bilateral minimum distance  at least $d$.
\end{lemma}
{\bf Proof:}
Since ${\vec 1} \in D$, 
we have $y+{\vec 1} \in D$, for each  $y\in D$, 
hence  $n - |y| = |y+1| \geq d$ if $y \neq {\vec 1}$. 
Thus,  if $C \subset D$ is an $\F_2$-linear code such that  ${\vec 1} \not\in C$, 
then the bilateral minimum distance of $C$ is at least $d$. Let $C$ be the set of even weight weight codewords of $D$.
Thus, ${\vec 1} \not\in C$ since $n$ is odd, hence  the bilateral minimum distance of $C$ is at least $d$
\footnote{
Note also that if $d$ is odd, which is the case in the examples below,  
then the bilateral minimum distance of $C$ is at least $d+1$.  Nevertheless, we will use the lower  bound $d$ on the 
bilateral minimum distance since  Theorem \ref{cr1} and Corollary \ref{main} assume that $d$ is odd. 
}.
The dual of $C$ is the $\F_2$-linear code generated by $D^\bot$ and ${\vec 1}$, i.e., 
$C^\bot= D^\bot \cup (D^\bot + {\vec 1})$. 
\finito 

\subsection{Extended Hadamard code}
Let $n=2^r-1$, where $r\geq 2$ is an integer, and let 
$D$ be the $(2^r-1,2^r-1-r,3)$-Hamming code.  Thus, $D^\bot$ is the $(2^r-1,r,2^{r-1})$-Hadamard code, 
and $Q= D^\bot \cup (D^\bot + {\vec 1})$ is the extended Hadamard code of size $2^{r+1}=2(n+1)$.
The all-ones vector ${\vec 1}\in D$ since all the codewords of the Hadamard code $D^\bot$ have even weight (the weight is either 
$0$ or $\frac{n+1}{2}=2^{r-1}$).   Thus, by  Lemma \ref{biconst}, the dual of $Q$ has bilateral minimum distance  at least $d=3$.
The weight distribution of $Q$ 
is given by 
\[
   \overline{\mu_Q}(w)=\left\{\begin{array}{ll} \frac{n}{2(n+1)}=\frac{1}{2}-O(\frac{1}{n})
 & \mbox{ if  } w = \frac{n-1}{2} \mbox{ or } \frac{n+1}{2}  \\ 
               \frac{1}{2(n+1)} = O(\frac{1}{n}) & \mbox{ if } w=0 \mbox{ or } n\\
               0 & \mbox{ otherwise.}\\
\end{array}\right.
\]
Thus, $\|m_Q-\bind_n\|_\infty = \Theta(1)$
and $\|m_Q-\bind_n\|_1 = \Theta(1)$. However, by Part (a) of  Theorem \ref{cr1} with $t=1$, we have 
\[
E_{u\sim U_n} \| m_{Q+u} - \bind_n \|_\infty  \leq  (\sqrt{2}e) \frac{\ln{(\frac{n}{2})}}{\sqrt{n}}=O\left( \frac{\ln{n}}{\sqrt{n}}\right).
\]
Note that Corollary \ref{maincor} is not useful here since it it gives nontrivial bounds for $t\geq 3$.

\subsection{Extended dual BCH code} 
Let $\F_{2^r}$ be the finite field with $2^r$ elements and 
 $\F_{2^r}^\times$ be the set of nonzero elements of $\F_{2^r}$.
If $r\geq  2$ and $t\geq 1$ are integers such that $2t-2< 2^{r/2}$, let 
$n = 2^r -1$, and consider the BCH code  $BCH(t,r) \subset\F_2^n$:
\[
      BCH(t,r) = \{ (f(a))_{a\in \F_{2^r}^\times} ~:~ f \in \F_{2^r}[x] \mbox{ s.t. } deg(f) \leq 2^r - 2t -1 \} \cap \F_2^{\F_{2^r}^\times}.
\]
We have (see \cite{MS77}):
\begin{itemize}
\item[a)] $dim(BCH(t,r)) = 2^r - 1 - rt$
\item[b)] The minimum distance  of $BCH(t,r)$ is at least $2t+1$
\item[c)](Weil-Carlitz-Uchiyama Bound) For each non-zero codeword $x\in BCH(t,r)^\bot$,  we have $||x|-2^{r-1}|\leq (t-1)2^{r/2}$, hence 
$||x|-(n+1)/2| \leq (t-1)\sqrt{n+1}$. 
\end{itemize}
Note that  the condition $2t-2< 2^{r/2}$ is equivalent to $t < \frac{1}{2}\sqrt{n+1}+1$.  
Let $D = BCH(r,t)$ and note that ${\vec 1}\in D$  (for $f =1$). 
Consider the dual BCH code  $D^\bot \subset \F_2^n$ and note that $|D^\bot|  = 2^{rt}=(n+1)^t$.
Then  $Q= D^\bot \cup (D^\bot + {\vec 1})$ is the extended dual BCH code of size $2(n+1)^t$.
By  Lemma \ref{biconst}, the dual of $Q$ has bilateral minimum distance  at least $d=2t+1$. 
\begin{lemma}\label{lembchdd}
If $t = \Theta(1)$, then $\| \overline{\mu_{Q}} - \bind_n \|_1 = \Theta(1)$.
\end{lemma}
However, it follows from Part (a) of Corollary \ref{maincor} that 
\[
   E_{u\sim U_n} \| \overline{\mu_{Q+u}} - \bind_n \|_1 
\leq  (2t+1)\left(e\ln{\frac{n}{2t}}\right)^{t}\left(\frac{2t}{n}\right)^{\frac{t}{2}-1}  = O\left(\frac{\ln^{t}{n}}{n^{\frac{t}{2}-1}}\right)
\]
for $t=O(1)$.  For $t\geq 3$ (i.e., $d\geq 7$), the decay  bypasses the $\Theta(1)$ error floor in Lemma \ref{lembchdd}. 

~\\
\noindent 
{\bf Proof of Lemma \ref{lembchdd}:}
By the Weil-Carlitz-Uchiyama Bound, we have  $\overline{\mu_Q}(w) = 0$ if $w\neq 0,n$ and $|w-(n+1)/2|> (t+1)\sqrt{n+1}$. Thus,  
\begin{eqnarray*}
\| \overline{\mu_{Q}} - \bind_n \|_1  &\geq& \bind_n(w\neq 0,n:|w-\mbox{$\frac{n+1}{2}$}|>(t+1)\sqrt{n+1}) \\
                                   &\geq& \bind_n(w:\mbox{$  \frac{n+1}{2}+(t+1)\sqrt{n+1} \leq w 
<  \frac{n+1}{2}+(t+2)\sqrt{n+1}$} ) \\
&&~~~\mbox{(since $\frac{n+1}{2}+(t+2)\sqrt{n+1} <n$, for $n$ large enough)}\\
                                   &\geq & \sqrt{n+1}~\bind_n(\mbox{$\frac{n+1}{2}$} + (t+2)\sqrt{n+1}) \\
                                   &\geq & \sqrt{\frac{2(n+1)}{\pi n}}e^{-2(t+2)^2}(1- o(1)), 
\end{eqnarray*}
using  de Moivre-Laplace normal approximation of the binomial 
 $\bind_n(w) = \sqrt{\frac{2}{\pi n}}e^{-2\frac{(w - n/2)^2}{n}}(1\pm o(1))$, 
                              which holds if $|w-n/2|=o(n^{2/3})$ (e.g., \cite{Fel68} page 184).   
\finito
\subsection{Conjecture}\label{conjs}
It follows from the  Extended dual BCH code example that 
polynomial-size codes 
may have a coset whose 
weight distribution is bounded away from the  binomial  distribution
 in the $L_1$-norm  by a constant error floor,  even 
if the code has a large dual bilateral distance. We believe that this error floor  is 
not due to the weakness of the ``large dual bilateral distance'' requirement, but it is simply 
due to the small-size and the linearity of the code.

\begin{conjecture}\label{conj1} 
For each constant $t>0$, there is a constant $\e>0$ such that, for $n$ large enough, for each $\F_2$-linear code  $Q\subset \F_2^n$ 
of size at most $n^t$, there exists a coset $u+Q$ of $Q$ for some $u \in \F_2^n$ such that 
$\| \overline{\mu_{Q+u}} - \bind_n\|_1 \geq \e$. 
\end{conjecture}
That is, polynomial-size linear codes do not behave like arbitrary random subsets of $\{0,1\}^n$ of the same size. 
We leave the question of proving or disproving the conjecture open. 

\section{Fourier transform preliminaries}\label{frprelS}

The study of error correcting codes  using  using  Harmonic analysis  methods dates back to  MacWillimas \cite{Mac63} 
(see also \cite{LMN93} for and the references therein).
We give below  some preliminary notions used in the proof. 

Identify the hypercube $\{0,1\}^n$ with the 
group  $\Z_2^n = (\Z/2\Z)^n$.
The   {\em characters}  of the abelian group $\Z_2^n$
are  $\{ \X_z \}_{z\in \Z_2^n}$,
 where  $\X_z:\lbrace 0,1 \rbrace^n \rightarrow \{-1,1\}$ is given by $\X_y(x) =  (-1)^{\langle x,y \rangle}$, and 
$\langle x, y\rangle = \sum_{i=1}^{n} x_iy_i$.

Consider the 
$\cmp$-vector space $\L(\Z_2^n)$  of complex valued functions $\Z_2^n\rightarrow \cmp$ 
endowed with the inner product $\langle, \rangle$ associated with the uniform distribution on $\{0,1\}^n$:
 $$\langle f,g\rangle = E_{U_n} f \overline{g}= \frac{1}{2^n} \sum_{x} f(x) \overline{g(x)},$$ 
 where $\bar{~}$ is the complex conjugation operator.   

 The characters $\{ \X_z \}_z$ form an orthonormal  basis of $\L(\Z_2^n)$, i.e., for each $z,z'\in \{0,1\}^n$, 
\[
  \langle \X_z,\X_{z'}\rangle     =\left\{\begin{array}{ll} 1 & \mbox{ if  } z=z'\\ 0 & \mbox{if } z\neq z'.\end{array}\right.
\]
If $f\in \L(\Z_2^n)$, its Fourier transform 
$\widehat{f}\in \L(\Z_2^n)$ is given by the coefficients of the unique 
 expansion of $f$ in terms of $\{\X_z\}_z$: \[
f(x) = \sum_{z} \widehat{f}(z)\X_z(x) 
\mbox{ }\mbox{ }\mbox{ and } \mbox{ }\mbox{ }
\widehat{f}(z) = \langle f, \X_z\rangle  =E_{U_n} f \X_z.
\]
Note that $\widehat{\widehat f} = 2^n f$.  

The {\em degree}  of $f$ 
 is the smallest 
degree of a polynomial  $p\in \cmp[x_1,\ldots,x_n]$ such that 
$p(x) = f(x)$ for all $x\in \{0,1\}^n$.
Equivalently, in terms of the Fourier transform $\widehat{f}$ over $\cmp$,
the degree of $f$ is equal to the maximal weight of 
 $z\in \Z_2^n$ such that  $\widehat{f}(z) \neq 0$.

 If   $f,g\in \L(\Z_2^n)$, we have   
\begin{equation}\label{inneq}  
\langle f, g\rangle = 2^n \langle \widehat{f}, \widehat{g}\rangle = \sum_{z} \widehat{f}(z) \widehat{g}(z). 
\end{equation}

{\bf Parseval's equality.}
A special case of (\ref{inneq}) is  {\em Parseval's equality}: 
\begin{equation}\label{pare}  
E_{U_n} |f|^2 = \sum_{z} |\widehat{f}(z)|^2 = \| \widehat{f}\|_2^2. 
\end{equation}
We need the following basic lemma which follows from Parseval's equality.
\begin{lemma}\label{parel}  
Let $\mu$ be a  probability distribution on $\{0,1\}^n$. For each    $u\in \{0,1\}^n$, define the translation $($mod $2)$
$\sigma_u \mu$ of $\mu$ by $u$ to be the 
probability distribution on  $\{0,1\}^n$ given by $(\sigma_u \mu)(x) = \mu(x+u)$. 
If $f:\{0,1\}^n\rightarrow \CN$, then 
\[ 
   E_{u\sim U_n} | E_{\sigma_u \mu} f-E_{U_n}f |^2 = \sum_{z\neq 0 } |\widehat{f}(z)|^2 (E_{\mu} \X_z)^2. 
\] 
\end{lemma}
{\bf Proof:} Define $\Delta:\{0,1\}^n\rightarrow \R$ by $\Delta(u)=E_{\sigma_u \mu} f-E_{U_n}f$.  
Consider the Fourier expansion of $f$:
$f
 = \sum_{z} \widehat{f}(z) \X_z$.
Thus, 
\[
   \Delta(u) = E_{y\sim \mu} \sum_{z} \widehat{f}(z) \X_z(y+u) - E_{U_n} f = 
 \sum_{z} \X_z(u) \widehat{f}(z) E_{\mu} \X_z  - E_{U_n}f = 
 \sum_{z\neq 0} \X_z(u) \widehat{f}(z) E_{\mu} \X_z 
\]
since  
 $\X_0 = 1$ and $ \widehat{f}(0)=E_{U_n} f$. Hence  
$\widehat{\Delta}(0) = 0$ and 
$\widehat{\Delta}(z) = \widehat{f}(z) E_{\mu} \X_z$ for each $z\neq 0$.  
The lemma then follows from Parseval's equality.  
\finito

{\bf Fourier transform of the exponential function.} The Fourier transform of the  exponential function is another  exponential function. 
\begin{lemma}\label{frprdcmp}
 Let $r$ be complex number and  
$g_r:\{0,1\}^n\rightarrow \CN$ be given by $g_r(x) = r^{|x|}$.
Then 
$
     \widehat{g_r}(z) =\left( \frac{1+r}{2}\right)^n \left(\frac{1-r}{1+r}\right)^{|z|}.
$ 
Moreover, if $r= e^{i\theta}$, then $\widehat{g_{r}}(z) =e^{i n\theta/2} \left(\cos\frac{\theta}{2}\right)^{n} \left(-i \tan{\frac{\theta}{2}}\right)^{|z|}$.
\end{lemma}
{\bf Proof:}
\begin{eqnarray*}
\widehat{g_r}(z) &=& \frac{1}{2^n}\sum_{x}  r^{\sum_i x_i}\X_z(x) 
               = \frac{1}{2^n}\sum_{x} \prod_{i=1}^n (r(-1)^{z_i})^{x_i} \\
               &=&  \prod_{i=1}^n \left(\frac{1+r(-1)^{z_i}}{2}\right) 
               = \left( \frac{1-r}{2}\right)^{|z|} \left( \frac{1+r}{2}\right)^{n-|z|}. 
\end{eqnarray*}
If $r = e^{i \theta}$, then $\frac{1+e^{i\theta}}{2}  = e^{i\theta/2}\cos\frac{\theta}{2}$ and 
 $\frac{1-e^{i \theta }}{1+e^{i \theta }}=\frac{e^{-i \theta/2 }-e^{i \theta/2 }}{e^{-i \theta/2 }+e^{i \theta/2 }} = 
-i \tan{\frac{\theta}{2}}$. 
\finito

{\bf Fourier transform and linear codes.} 
If $Q\subset \F_2^n$  is an $\F_2$-linear code and $Q^\bot$ is its dual, then for each $z\in \F_2^n$, we have 
\begin{equation}\label{frdl}
  \sum_{y\in Q} \X_z(y)    =\left\{\begin{array}{ll} |Q| & \mbox{ if  } z\in Q^\bot\\ 0 & \mbox{ otherwise. }\end{array}\right.
\end{equation}
It follows that 
\begin{equation}\label{frd2}
  E_{\mu_Q} \X_z    =\left\{\begin{array}{ll} 1 & \mbox{ if  } z\in Q^\bot\\ 0 & \mbox{ otherwise. }\end{array}\right.
\end{equation}
Finally, we need MacWilliams's identity:
\begin{lemma}[MacWilliams's identity\cite{Mac63}]\label{MacId} 
Let $Q\subset \F_2^n$ be an $\F_2$-linear code and $s$ be a complex number, 
 then $\sum_{z\in Q^\bot} s^{|z|} = (1+s)^n E_{x\in Q} \left(\frac{1-s}{1+s}\right)^{|x|}$.
\end{lemma}
Since the proof is direct  given the above machinery, we add it for completeness.
~\\
{\bf Proof:}  Let $r = \frac{1-s}{1+s}$. By Lemma \ref{frprdcmp},  
$g_r(x) = \sum_{z} \left( \frac{1+r}{2}\right)^n \left(\frac{1-r}{1+r}\right)^{|z|} \X_z(x)$.
It follows from (\ref{frd2}) that $E_{x\in Q} g_r(x) =\left( \frac{1+r}{2}\right)^n  \sum_{z\in Q^\bot} \left(\frac{1-r}{1+r}\right)^{|z|}$. 
MacWilliams's identity thus follows from the relations $\frac{1-r}{1+r}=s$ and $\frac{1+r}{2}=\frac{1}{1+s}$.  
\finito

\section{Proof of Theorem \ref{main} }\label{ProofS} 
\begin{definition}
If $0\leq \theta < 2\pi$, define $e_\theta:\{0,1\}^n \rightarrow {\Bbb C}$ by $e_\theta(x) \defeq e^{i\theta |x|}$. 
If $Q\subset \F_2^n$ is an $\F_2$-linear code, 
 define $\Delta_{Q,\theta}:\{0,1\}^n \rightarrow {\Bbb C}$  by 
$\Delta_{Q,\theta}(u)\defeq E_{\mu_{Q+u}} e_\theta - E_{U_n} e_{\theta}$. 
\end{definition} 
We restate below  Theorem \ref{main} in terms of $\Delta_{Q,\theta}$.

~\\
\noindent
{\bf Theorem~\ref{main}} {\bf (Mean-square-error bound)} {\em  
Let $Q\subsetneq \F_2^n$  be an $\F_2$-linear code whose dual $Q^\bot$ has bilateral minimum distance at least $d=2t+1$, where $t\geq 1$ is an integer.
Then, for each $0\leq \theta < 2\pi$, we have the  bounds: 
\begin{itemize}
\item[a)]  {\em $($Small dual distance bound$)$} 
\[
      E_{U_n} |\Delta_{Q,\theta}|^2  \leq \left(e\ln{\frac{n}{2t}}\right)^{2t}\left(\frac{2t}{n}\right)^{t} 
\]
\item[b)]  {\em $($Large dual distance bound$)$} 
\[
      E_{U_n} |\Delta_{Q,\theta}|^2 \leq 2 e^{-\frac{t}{5}}.
\]
\end{itemize}
}

~\\
At a high level, our proof technique is as follows.  
First we show in Lemma \ref{mainl1} that 
$$  E_{U_n} |\Delta_{Q,\theta}|^2
 = E_{w\sim \overline{\mu_Q}} c^w - 
\left(\frac{c+1}{2}\right)^n,$$
where $c = \cos{\theta}$.
Note that 
$\left(\frac{c+1}{2}\right)^n = E_{w\sim \bind_n} c^w$.
Lemma \ref{mainl1} exploits the linearity of the code in subtle manner. 
The starting to point is to express 
$E_{U_n} |\Delta_{Q,\theta}|^2$ in terms of the  squared norm 
of the Fourier coefficients of $e_\theta$ and the dual of $Q$ 
using Lemma \ref{parel} applied to $f=e_\theta$. We establish 
 Lemma \ref{mainl1} using the expression of the Fourier transform of $e_\theta$ in 
Lemma \ref{frprdcmp} and  using  MacWilliams's identity to go back to the original domain.  
The argument seems convoluted since after going the Fourier domain, we use 
MacWilliams's identity to go back to the original domain.  
The catch is that  MacWilliams's identity is used after Parseval's equality (Lemma \ref{parel} is based on  Parseval's equality)
which involves 
  squaring the  norm of the Fourier coefficients of 
$e_\theta$.
The  claim  can be established without going to the Fourier domain by algebraically  exploiting the code linearity  
but the proof is more complicated.

Then we bound 
$E_{w\sim \overline{\mu_Q}} c^w- \left(\frac{c+1}{2}\right)^n$ by ignoring the code linearity, 
and maximizing over the choice of probability distributions 
$\mu$ on $\{0,1\}^n$ such that $E_\mu \X_z = 0$ 
for all nonzero $z\in \{0,1\}^n$ such that $|z|\leq d-1$ 
or $|z|\geq n-d+1$. This property is satisfied by $\mu_Q$ as $Q^\bot$ has bilateral 
minimum distance at least $d$ (see (\ref{frd2})).  The constraints on $\mu$ define 
a linear program. Due to the symmetry of the problem, we note in Lemma \ref{mainl2}, that  
the dual linear program is 
$\min_{h} E_{\bind_n} h$, 
where we are optimizing on the all functions $h:[0:n]\rightarrow \R$ such that:  
\begin{itemize}
\item[i)]
$h$ can be expressed 
as $h(w) = f(w) + (-1)^w g(w)$, for some polynomials  $f(x),g(x) \in \R[x]$ each degree at most $d-1$
\item[ii)]
$h(w) \geq  c^w - \left(\frac{c+1}{2}\right)^n$, for all $w\in [0:n]$.
\end{itemize}   
That is $E_{U_n} |\Delta_{Q,\theta}|^2\leq \min_{h} E_{\bind_n} h$. 
We construct $h$ in Lemma \ref{mainl3}.  Let $k=d-1=2t$. 
We use three different constructions of $h$ depending on whether 
$c^* \leq c \leq 1$, $-1 \leq c \leq -c^*$, or   $|c|<c^*$, where $c^*$ is a parameter which will optimized. 
If $c^* \leq c \leq 1$, we  construct $h$ by truncating the Taylor series expansion of $c^x$ around $n/2$ to obtain a polynomial of degree $k$. 
If $-1< c \leq -c^*$, we write $c^w = (-1)^w |c|^w$ and we suitably 
construct $h$  using $(-1)^w$ and the first $k$ terms of the Taylor series expansion of $|c|^x$ around $n/2$.
If $|c|<c^*$, we set $h$ to a degree $k$ polynomial of the form $h(w)=a(n/2-w)^k + b$, where 
$a,b>0$ are suitably chosen parameters.  The constructions rely on the fact that $k$ is even.

\begin{lemma}\label{mainl1}
Let $Q\subsetneq \F_2^n$  be an $\F_2$-linear code and $0\leq \theta < 2\pi$. 
Then 
\[
   E_{U_n} |\Delta_{Q,\theta}|^2
 = E_{w\sim \overline{\mu_Q}} (\cos{\theta})^w - \left(\frac{\cos{\theta}+1}{2}\right)^n.
\]
\end{lemma}
Note that if  $\cos{\theta}
 = 0$ and $w=0$, then $0^0 =1$ is interpreted as  the limit of 
$(\cos{\theta})^0$ as  $\cos{\theta}$ tends to $0$. 

\noindent 
{\bf Proof:} 
Applying Lemma \ref{parel} to $f = e_\theta$, we get  
\[ 
   E_{U_n} |\Delta_{Q, \theta}|^2 = \sum_{z\neq 0 } |\widehat{e_\theta}(z)|^2 (E_{\mu_Q} \X_z)^2. 
\] 
We know from (\ref{frd2}) that 
\[
  E_{\mu_Q} \X_z    =\left\{\begin{array}{ll} 1 & \mbox{ if  } z\in Q^\bot\\ 0 & \mbox{ otherwise. }\end{array}\right.
\]
 It follows that 
\begin{equation} \label{gfr}
  E_{U_n} |\Delta_{Q, \theta}|^2 = \sum_{z\neq 0 \in Q^\bot} |\widehat{e_\theta}(z)|^2
 = \sum_{z\in Q^\bot} |\widehat{e_\theta}(z)|^2  - |\widehat{e_\theta}(0)|^2. 
\end{equation} 
By Lemma \ref{frprdcmp} applied with $r= e^{i\theta}$, we have 
$\widehat{e_\theta}(z) =e^{i n\theta/2} \left(\cos\frac{\theta}{2}\right)^{n} \left(-i \tan{\frac{\theta}{2}}\right)^{|z|}$.
Thus, 
\[
\sum_{z\in Q^\bot} |\widehat{e_\theta}(z)|^2 =\left(\mbox{$\cos\frac{\theta}{2}$}\right)^{2n} \sum_{z\in Q^\bot} \left(\mbox{$\left(\tan{\frac{\theta}{2}}\right)^2$}\right)^{|z|}.
\]
Applying MacWilliams's identity (Lemma \ref{MacId}) with $s  = \left(\tan{\frac{\theta}{2}}\right)^2$, we obtain 
\begin{equation}\label{gfr2}
\sum_{z\in Q^\bot} |\widehat{e_\theta}(z)|^2 =
\left(\mbox{$\cos{\frac{\theta}{2}}$}\right)^{2n} \left(1+\mbox{$\left(\tan{\frac{\theta}{2}}\right)^2$}\right)^n E_{x\in Q} \left(\frac{1-\left(\tan{\frac{\theta}{2}}\right)^2 }{1+\left(\tan{\frac{\theta}{2}}\right)^2}\right)^{|x|}
= E_{x\in Q} \left(\cos{\theta}\right)^{|x|}, 
\end{equation}
since  
$(\cos{\frac{\theta}{2}})^2(1+(\tan{\frac{\theta}{2}})^2)  = (\cos{\frac{\theta}{2}})^2+(\sin{\frac{\theta}{2}})^2=1$ and 
$\frac{1-\left(\tan{\frac{\theta}{2}}\right)^2 }{1+\left(\tan{\frac{\theta}{2}}\right)^2} =  \left(\sin{\frac{\theta}{2}}\right)^2-\left(\cos{\frac{\theta}{2}}\right)^2 =  \cos{\theta}$.   

Finally, applying  Lemma \ref{frprdcmp} again with $r= e^{i\theta}$ and $z=0$, we get 
$|\widehat{e_\theta}(0)|^2=\left(\cos\frac{\theta}{2}\right)^{2n}= \left(\frac{\cos{\theta}+1}{2}\right)^n$.
Replacing  with  (\ref{gfr2}) in (\ref{gfr}), we obtain 
\[
   E_{U_n} |\Delta_{Q, \theta}|^2  
= \sum_{w=0}^n \frac{\#\{y\in Q: |y|=w\}  }{|Q|} (\cos{\theta})^w
 - \left(\frac{\cos{\theta}+1}{2}\right)^n.
\]
\finito

\begin{lemma}\label{polyd}
Let $Q\subsetneq \F_2^n$  be an $\F_2$-linear code whose dual $Q^\bot$ has bilateral minimum distance at least $d\geq 1$. 
\begin{itemize}
\item[a)]   If $r:\{0,1\}^n\rightarrow \R$ can be expressed as 
  $r(x) = p(x) + (-1)^{|x|}q(x)$  for some  
$p,q:\{0,1\}^n\rightarrow \R$ each of degree (see section \ref{frprelS}) at most $d-1$,  
then  $E_{\mu_Q} r = E_{U_n} r$.
\item[b)] If $h:[0:n]\rightarrow \R$ can be expressed as 
 $h(w) = f(w) + (-1)^w g(w)$ for some polynomials  $f(x),g(x) \in \R[x]$ each degree at most $d-1$, 
then $E_{\overline{\mu_Q}} h = E_{\bind_n} h$.  
\end{itemize}
\end{lemma} 
{\bf Proof:}
\begin{itemize}
\item[(a)]
By (\ref{frd2}),  
 we have
\[
  E_{\mu_Q} \X_z    =\left\{\begin{array}{ll} 1 & \mbox{ if  } z\in Q^\bot\\ 0 & \mbox{ otherwise }\end{array}\right.
\]
Since $Q^\bot$ has bilateral minimum distance at least $d$, then $d\leq |z|\leq n-d$, for each nonzero $z\in Q^\bot$. Thus,  
  $E_{\mu_Q} \X_z = 0$ for each nonzero $z\in \{0,1\}^n$ such that $|z|\leq d-1$ or $|z|\geq n-d+1$. 

Consider the Fourier expansions $p(x) = \sum_{z:|z|\leq d-1} \widehat{p}(z) \X_z(x)$ and 
$q(x) = \sum_{z:|z|\leq d-1} \widehat{q}(z) \X_z(x)$. Since $(-1)^{|x|} \X_z(x)=\X_{z+\vec{1}}(x)$ 
and $|z+\vec{1}|=n-|z|$, where $\vec{1} \in \{0,1\}^n$  is the all-ones vector, we obtain 
\[
  r(x) = \sum_{z:|z|\leq d-1} \widehat{q}(z) \X_z(x)+\sum_{z:|z|\geq n- d+1} \widehat{r}(z+\vec{1}) \X_z(x)
\]
It follows that $E_{\mu_Q} r = \widehat{q}(0) = E_{U_n} r$.
\item[(b)]  Let 
$p,q,r:\{0,1\}^n\rightarrow \R$ be given by 
 $p(x)=f(|x|)$, $q(x)=g(|x|)$, and $r(x) = p(x) + (-1)^{|x|}q(x)$. 
Thus, $p$ and $q$ are each of degree at most $d-1$ and 
$r(x) = h(|x|)$. It follows from (a) that 
$E_{\mu_Q} r = E_{U_n} r$. Since $r$ is a symmetric function (i.e., $r(x)$ depends only on the weight $|x|$ of $x$), we have 
$E_{\mu_Q} r = E_{\overline{\mu_Q}} h$ 
and $E_{U_n} r = E_{\bind_n} h$. It follows that   $E_{\overline{\mu_Q}} h = E_{\bind_n} h$.  \finito
\end{itemize}

\begin{definition}
If  $-1\leq c \leq 1$, define $H_c^{(n)}:[0:n]\rightarrow \R$ by 
\[
   H_c^{(n)}(w) \defeq c^w - \left(\frac{c+1}{2}\right)^n.
\]
Here again, if  $c = 0$ and $w=0$, then $0^0 =1$ is interpreted as  the limit of $c^0$ as  $c$ tends to $0$. 
\end{definition}
Note that $E_{w\sim \bind_n} c^w =  \left(\frac{c+1}{2}\right)^n$.
\begin{lemma}[LP duality]\label{mainl2}
  Let $Q\subsetneq \F_2^n$  be an $\F_2$-linear code whose dual $Q^\bot$ has bilateral minimum distance at least $d\geq 1$, and  let $0\leq \theta < 2\pi$. 
Then 
\[
   E_{U_n} |\Delta_{Q,\theta}|^2 \leq \min_{h} E_{\bind_n} h,   
\]
where we are optimizing on the all functions $h:[0:n]\rightarrow \R$ such that: 
\begin{itemize}
\item[i)]
$h$ can be expressed 
as $h(w) = f(w) + (-1)^w g(w)$, for some polynomials  $f(x),g(x) \in \R[x]$ each degree at most $d-1$
\item[ii)]
$h(w) \geq H_{\cos{\theta}}^{(n)}(w)$ for all $w\in [0:n]$. 
\end{itemize}
\end{lemma} 
{\bf Proof:} 
Using Lemma \ref{mainl1}, $E_{U_n} |\Delta_{Q,\theta}|^2=
 E_{\overline{\mu_Q}} H_{\cos{\theta}}^{(n)}$.
Since $h \geq H_{\cos{\theta}}$, we have 
  $E_{\overline{\mu_Q}} H_{\cos{\theta}}^{(n)} \leq          E_{\overline{\mu_Q}} h$.
By  Lemma  \ref{polyd}, 
     $E_{\overline{\mu_Q}} h = E_{\bind_n} h$. It follows that 
$E |\Delta_{Q,\theta}|^2 \leq  E_{\bind_n} h$.

Note that the above argument is  weak LP duality. The converse 
also holds, in the sense that it not hard to verify that
the following linear programs are duals of each others: 
 \[
  \min_{h} E_{\bind_n} h = \max_{\gamma} E_\gamma H_{\cos{\theta}}^{(n)} 
\]
where the min is over all functions  $h:[0:n]\rightarrow \R$ satisfying (i) and (ii), and 
the max is over all probability distributions $\gamma$ on $[0:n]$ such that 
$E_\gamma h = E_{\bind_n} h$ for each $h$ satisfying (i).  
\finito

Theorem \ref{main} follows from Lemma \ref{mainl2} and  
Lemma \ref{mainl3} below. 
Note that Lemma \ref{mainl3} assumes that $n \geq 2d$, which must be the case by the definition of 
bilateral minimum distance.

\begin{lemma}[Construction] \label{mainl3}
Let $n$ and $t$ be integers such that $t\geq 1$ and $n\geq 2d$, where $d = 2t+1$. 
\begin{itemize}
\item[a)] For  all $-1\leq c \leq 1$, 
there exist polynomials $f(x),g(x) \in \R[x]$, each degree at most $d-1$, 
 such that with $h(w) = f(w) + (-1)^w g(w)$, we have  $h(w) \geq H_c^{(n)}(w)$ for all $w\in [0:n]$, and 
\[
E_{\bind_n}  h  \leq \left(e\ln{\frac{n}{2t}}\right)^{2t}\left(\frac{2t}{n}\right)^{t}.  
\]
\item[b)] For  all $-1\leq c \leq 1$, 
there exist polynomials $f(x),g(x) \in \R[x]$, each degree at most $d-1$, 
 such that with $h(w) = f(w) + (-1)^w g(w)$, we have  $h(w) \geq H_c^{(n)}(w)$ for all $w\in [0:n]$, and 
\[
E_{\bind_n}  h  \leq  2 e^{-\frac{t}{5}}.
\]
\end{itemize}
\end{lemma} 
{\bf Proof:}  
Let $k=d-1 = 2t$,  thus $k$ is even  and $k\geq 2$.  
Let $c^*= e^{-2\frac{k}{n}\beta}$, where $\beta>0$ is a parameter which we will optimize later. 
We use three different constructions depending on the value  of $c$: 
$c^* \leq c \leq 1$, 
$-1 \leq c \leq -c^*$, 
and $|c|<c^*$. 
At the end of the proof, we will set $\beta = \ln{\frac{n}{k}}$ to establish (a), and we will 
set $\beta$ to a constant to establish (b).

~\\\noindent
\underline{\bf Case 1:}  Assume that $c^* \leq c \leq 1$. 
Consider the Taylor approximation of the exponential around $0$:
\[
   e^x = \sum_{i=0}^{k-1} \frac{x^i}{i!} + e^{\alpha}\frac{x^{k}}{k!}
\]
for some number $\alpha$ between $0$ and $x$. We will use it to approximate 
$c^w$ around $n/2$:
\begin{eqnarray}
   c^w &=& c^{n/2} c^{w-n/2} = c^{n/2} e^{(n/2-w)\ln\frac{1}{c}}  \nonumber \\
       &=& a_c(w)+ c^{n/2}e^{\alpha_c(w)}  b_c(w),\label{tayapp}
\end{eqnarray}
where: 
\begin{itemize}
\item[i)] $a_c(w) = c^{n/2}\sum_{i=0}^{k-1} \frac{1}{i!}\left(\left(\frac{n}{2}-w\right)\ln\frac{1}{c}\right)^i$
\item[ii)] $b_c(w) =  \frac{1}{k!}\left(\left(\frac{n}{2}-w\right)\ln\frac{1}{c}\right)^{k}$
\item[iii)] $\alpha_c(w)$ is a number between $0$ and $(\frac{n}{2}-w)\ln\frac{1}{c}$. 

Thus, for all $w\in [0:n]$, we have  $\alpha_c(w) \leq \frac{n}{2}\ln\frac{1}{c}$, and hence 
$ c^{n/2}e^{\alpha_c(w)} \leq 1$. Moreover,  $c^{n/2}e^{\alpha_c(w)} \geq 0$. Thus, 
$0 \leq  c^{n/2}e^{\alpha_c(w)} \leq 1$. 
\end{itemize}
Since $k$ is even, $b_c(w)\geq 0$ for all $w$. 
  Therefore, $c^{n/2}e^{\alpha_c(w)}  b_c(w)\leq b_c(w)$ for all $w\in [0:n]$ 
(since by (iii),  $ c^{n/2}e^{\alpha_c(w)} \leq 1$). 
Accordingly, $c^w \leq a_c(w) + b_c(w)$, for all $w\in [0:n]$.
Let $$h(w) =  a_c(w) + b_c(w) -  \left(\frac{c+1}{2}\right)^n.$$ 
Thus,  $deg(h)= k= d-1$ and 
$h(w)\geq H_{c}(w)$. 

Since $E_{w\sim \bind_n} c^w =  \left(\frac{c+1}{2}\right)^n$, we have 
\begin{eqnarray}
E_{\bind_n} h 
&=& E_{w\sim \bind_n} a_c(w) + b_c(w)  - c^w\nonumber \\
&=&E_{w\sim \bind_n}(1-c^{n/2}e^{\alpha_c(w)})  b_c(w)\nonumber \\
&\leq & E_{w\sim \bind_n}  b_c(w)   ~~~\mbox{(since $c^{n/2}e^{\alpha_c(w)}\geq 0$)}\nonumber \\
&=&  \frac{1}{k!}\left(\frac{n}{2}\ln{\frac{1}{c}}\right)^kE_{w\sim \bind_n}\left(\frac{n/2-w}{n/2}\right)^k\nonumber \\
&\leq &\frac{1}{k!} (k \beta)^kE_{w\sim \bind_n}\left(\frac{n/2-w}{n/2}\right)^k ~~~\mbox{(since $c\geq  c^* =  e^{-2\frac{k}{n}\beta}$)}\nonumber 
\end{eqnarray}
Using Lemma \ref{binommom} below, we obtain 
\begin{eqnarray}
E_{\bind_n} h 
&\leq &  \frac{1}{k!} (k \beta)^k \left(\frac{k}{n}\right)^{k/2}\nonumber
\\
&\leq &  \frac{1}{2}(e\beta)^k\left(\frac{k}{n}\right)^{k/2} \label{c1e} 
\end{eqnarray}
by Stirling Approximation: $k!\geq \sqrt{2\pi} k^{k+1/2} e^{-k}\geq 2 k^{k} e^{-k}$, which holds for all for $k\geq 1$.

\begin{lemma}\label{binommom} For each even $k \geq 2$, we have 
\[
  E_{w\sim \bind_n}\left(\frac{n/2-w}{n/2}\right)^k
 \leq \left(\frac{k}{n}\right)^{k/2}.
\]
\end{lemma}
{\bf Proof:} We have
\begin{eqnarray*}
  E_{w\sim \bind_n}\left(\frac{n/2-w}{n/2}\right)^k 
&=&E_{x\sim U_n }\left(\frac{\sum_{i=1}^n (-1)^{x_i}}{n}\right)^k  ~~~\mbox{(since $\frac{n}{2}-|x| = \frac{1}{2}\sum_i (-1)^{x_i}$)}\\
&=&\frac{1}{n^k}  \sum_{i_1,\ldots, i_k\in [n]} E_{x\sim U_n} (-1)^{\sum_{t=1}^k x_{i_t}}=\frac{A_{n,k}}{n^k},  
\end{eqnarray*}
where $A_{n,k}$ is the number of $k$-tuples $(i_1,\ldots, i_k)\in [n]^k$ such that 
$(i_1,\ldots, i_k)$ can be partitioned into $k/2$ disjoint pairs of equal integers. 
Thus, $A_{n,k}\leq n(k-1) n(k-3) n(k-5) \ldots n \leq (n k)^{k/2}$.
 
\lemfinito 

~\\\noindent
\underline{\bf Case 2:}  Assume that  $-1 \leq c \leq -c^*$. Thus, $c^w = (-1)^w |c|^w$ and 
$c^* \leq |c| \leq 1$. We use the Taylor approximation of $|c|^w$ in Case 1 (Equation (\ref{tayapp})):
\[
  |c|^w =  a_{|c|}(w)+ |c|^{n/2}e^{\alpha_{|c|}(w)}  b_{|c|}(w),
\]
where $a_{|c|}$ and $b_{|c|}$ and $\alpha_{|c|}$ are as given in (i),(ii), and (iii) above with $|c|$ instead of $c$.  
Hence
\[
  c^w =  (-1)^wa_{|c|}(w)+ (-1)^w|c|^{n/2}e^{\alpha_{|c|}(w)}  b_{|c|}(w). 
\]
Let $$h(w) =  (-1)^w a_{|c|}(w) + b_{|c|}(w) -  \left(\frac{c+1}{2}\right)^n.$$

We have $deg(a_{|c|})=k-1< d-1$ and $deg(b_{|c|})=k= d-1$.

Since $k$ is even, $b_{|c|}(w)\geq 0$ for all $w$. 
We know from (iii) that 
$ 0\leq |c|^{n/2}e^{\alpha_{|c|}(w)} \leq 1$, for all $w\in [0:n]$.  
  Therefore, $(-1)^w|c|^{n/2}e^{\alpha_{|c|}(w)}  b_{|c|}(w)\leq b_{|c|}(w)$ for all $w\in [0:n]$.
Accordingly $c^w \leq (-1)^wa_{|c|}(w) + b_{|c|}(w)$, for all $w\in [0:n]$.
Therefore, $ H_{c}(w) \leq h(w)$. 

To bound $E_{\bind_n} h$, we proceed  as in Case 1.  The only difference is that we get an extra factor of $2$. 
We have  $E_{w\sim \bind_n} c^w =  
\left(\frac{c+1}{2}\right)^n$, thus 
\begin{eqnarray}
E_{\bind_n} h &=& E_{w\sim \bind_n} (-1)^wa_{|c|}(w) + b_{|c|}(w)  - c^w\nonumber\\
&=&E_{w\sim \bind_n}(1-(-1)^w|c|^{n/2}e^{\alpha_{|c|}(w)})  b_{|c|}(w)\nonumber\\
&\leq & 2 E_{w\sim \bind_n}  b_{|c|}(w)   ~~~\mbox{(since $0 \leq |c|^{n/2}e^{\alpha_{|c|}(w)}\leq 1$)}\nonumber\\
&=&  2\frac{1}{k!}\left(\frac{n}{2}\ln{\frac{1}{|c|}}\right)^kE_{w\sim \bind_n}\left(\frac{n/2-w}{n/2}\right)^k \nonumber\\
&\leq &2\frac{1}{k!} (k \beta)^k             E_{w\sim \bind_n}\left(\frac{n/2-w}{n/2}\right)^k                       ~~~\mbox{(since $|c|\geq  c^* =  e^{-2\frac{k}{n}\beta}$)}\nonumber \\ 
&\leq &  (e\beta)^k\left(\frac{k}{n}\right)^{k/2}    ~~~\mbox{(by arguing as in Case 1).} \label{c2e} 
\end{eqnarray}

~\\\noindent
\underline{\bf Case 3:} 
 Assume that $|c|  < c^*$. Then 
\[
   H_c^{(n)}(w) = c^w - \left(\frac{c+1}{2}\right)^n < c^w \leq {c^*}^w =  e^{-\frac{2kw\beta}{n}}. 
\]
Note that $c+1>0$ since $|c|<c^*<1$. 

Let $0<\gamma<1/2$ be a parameter which will be optimized. 
If $w\geq \gamma n$, we use the bound  $H_c^{(n)}(w) \leq  e^{-2\beta\gamma k}$.
If $w< \gamma n$, we use the bound    $H_c^{(n)}(w) \leq  1$. 
Let 
\[
   h(w) = \frac{1}{(1-2\gamma)^k} \left(\frac{n/2-w}{n/2}\right)^k + e^{-2\beta\gamma k}.
\]
Thus, if $w\geq \gamma n$,  we have  $H_c^{(n)}(w) \leq  e^{-2\beta\gamma k} \leq  h(w)$ (recall that $k$ is even, hence $(n/2-w)^k \geq 0$).
If $w<\gamma n$, i.e., $1-2\gamma<\frac{n/2-w}{n/2}$,  then
$1< \frac{1}{(1-2\gamma)^k} \left(\frac{n/2-w}{n/2}\right)^k$, hence $H_c^{(n)}(w) \leq 1 < h(w)$. 
It follows that $h$ is a degree $k= d-1$ polynomial such that $h \leq H_c^{(n)}$. Moreover, 
\begin{eqnarray}
  E_{\bind_n} h &=& \frac{1}{(1-2\gamma)^k} E_{w \sim \bind_n} \left(\frac{n/2-w}{n/2}\right)^k + e^{-2\beta\gamma k} \nonumber \\
&\leq& \frac{1}{(1-2\gamma)^k}   \left(\frac{k}{n}\right)^{k/2} + e^{-2\beta\gamma k} ~~~\mbox{(by Lemma \ref{binommom}).} \label{c3e} 
\end{eqnarray}

~\\
The bound in (\ref{c2e}) in Case 2 is twice that in (\ref{c1e})  in Case 1, hence we can focus on Cases 2 and 3, i.e., 
 (\ref{c2e})  and (\ref{c3e}).

First, we establish the bound in Part (a). Set $\beta = \ln{\frac{n}{k}}$ and $\gamma = \frac{1}{4}$, hence (\ref{c2e}) reduces to 
$\left(e\ln{\frac{n}{k}}\right)^k\left(\frac{k}{n}\right)^{k/2}$, and (\ref{c3e}) reduces to 
$2^k\left(\frac{k}{n}\right)^{k/2} + \left(\frac{k}{n}\right)^{k/2}\leq \left(e\ln{\frac{n}{k}}\right)^k\left(\frac{k}{n}\right)^{k/2}$ if 
$\frac{n}{k}\geq 4$. It follows that, if $\frac{n}{k}\geq 4$, then -- in each of the 3 cases -- 
there exists $h$ such that $$E_{\bind_n}  h  \leq  \left(e\ln{\frac{n}{k}}\right)^k\left(\frac{k}{n}\right)^{k/2} = 
\left(e\ln{\frac{n}{2t}}\right)^{2t}\left(\frac{2t}{n}\right)^{t}. $$
We can ignore the $\frac{n}{k}\geq 4$ assumption since 
the bound $(e\ln{\frac{n}{k}})^{k}(\frac{k}{n})^{k/2}
= ((e\ln{\frac{n}{k}})^2\frac{k}{n})^{k/2}
\geq 1$, for 
$\frac{n}{k}\leq 212$. 
Moreover, since $H_c^{(n)}(w)\leq 1$ for all $w$ and $c$, by setting $h=1$, we can trivially achieve $E_{\bind_n} h=1$.

To establish Part (b),  set $\beta = \frac{1}{e(1-2\gamma)}$ and $\gamma = 0.107$,
hence (\ref{c2e}) reduces to 
$\frac{1}{(1-2\gamma)^k}\left(\frac{k}{n}\right)^{k/2}$, thus (\ref{c3e}) is the larger bound. 
Since $k=d-1$ and $d\leq n/2$, we have $\frac{k}{n}\leq \frac{1}{2}$. It follows that (\ref{c3e}) is at most 
\[
   \frac{1}{(1-2\gamma)^k}   \left(\frac{1}{2}\right)^{k/2} + e^{-2\beta\gamma k}  = e^{-(0.1057..)k}+e^{-(0.1001..)k}\leq 2e^{-k/10} =  2e^{-t/5}.
\]

\finito

\section{Comparison with  small-bias probability distributions}\label{sbias}

In this section, we compare our $L_1$-bound  (Corollary \ref{maincor}) with the bound in Corollary 6.2 in \cite{Baz14},
  which is the analogue of our $L_1$-bound for small-bias probability distributions.   
Roughly speaking,  a probability distribution  on $\{0,1\}^n$ has small bias if it looks like the uniform distribution for 
 all parity functions on subsets of the $n$ input variables.
More formally, let $\delta\geq 0$. A probability distribution $\mu$ on $\{0,1\}^n$  
 is {\em $\delta$-biased}
if  $|E_\mu \X_z|  \leq \delta$ for each nonzero $z\in \{0,1\}^n$  \cite{NN93}.  
Recall that 
if $\mu$ is a  probability distribution on $\{0,1\}^n$  and $u\in \{0,1\}^n$, 
then the  $\F_2$-translation 
$\sigma_u \mu$ of $\mu$ by $u$ is the 
probability distribution on  $\{0,1\}^n$ given by $(\sigma_u \mu)(x) = \mu(x+u)$. 
The  bound in Corollary 6.2 in \cite{Baz14} is $E_{u\sim U_n} \| \overline{\sigma_u \mu} -\bind_n\|_1 \leq \delta \sqrt{n+1}$, 
i.e., the average 
$L_1$-distance between  the binomial distribution and the weight distribution  of the translation of $\mu$ by a random vector in $\{0,1\}^n$ 
is  at most $\delta \sqrt{n+1}$.  
The key behind this bound is following observation which  is inspired by the paper of Viola \cite{Vio08} 
(the argument used to establish Lemma 3 in \cite{Vio08}). 
\begin{lemma}[\cite{Baz14}, Lemma 6.1]\label{absb} 
If $f:\{0,1\}^n\rightarrow \CN$ and $\mu$ be a $\delta$-biased probability distribution on $\{0,1\}^n$, then 
\[ 
   E_{u\sim U_n} | E_{\sigma_u \mu} f-E_{U_n}f |^2 \leq \delta^2 (E_{U_n}|f|^2-|E_{U_n}f|^2). 
\] 
\end{lemma}
The proof  follows from Parseval's equality and the definition of small bias.  
For completeness, we derive it below using Lemma \ref{parel} (which is also based on  Parseval's equality). 
~\\
{\bf Proof:}  
By Lemma \ref{parel}, 
\[ 
   E_{u\sim U_n} | E_{\sigma_u \mu} f-E_{U_n}f |^2 = \sum_{z\neq 0 } |\widehat{f}(z)|^2 (E_{\mu} \X_z)^2 \leq \delta^2 \sum_{z\neq 0 } |\widehat{f}(z)|^2. 
\] 
We have $\widehat{f}(0) = E_{U_n} f$ and,  
by  Parseval's equality  (\ref{pare}),  
$E_{U_n} |f|^2 = \sum_{z} |\widehat{f}(z)|^2$.  It follows that $\sum_{z\neq 0 } |\widehat{f}(z)|^2 =E_{U_n}|f|^2-|E_{U_n}f|^2$.
\finito

Lemma \ref{absb}  is the analog of Theorem \ref{main} for small-bias spaces. 
The proof of  Lemma \ref{absb}  is significantly simpler. It is based on the fact $|E_\mu \X_z|  \leq \delta$ for each nonzero $z\in \{0,1\}^n$. 
In this context,  the key weakness of linear codes  is that $E_{\mu_Q}\X_z = 1$ for all each 
$z$ in the dual $Q^\bot$, which  is huge for small codes. 
On the other hand, the fact that $E_{\mu_Q}\X_z = 0$ for all $z\not\in Q^\bot$ is essential 
for the correctness Theorem \ref{main} in the sense that its proof breaks down 
if we ignore the  linearity of the code and focus on the 
constraint that 
$E_\mu \X_z  =0$ for each nonzero $z\in \{0,1\}^n$ of weight less than $d$ or larger than $n-d$. 
Finally, we note that Lemma \ref{absb} is more general than  Theorem \ref{main} 
as it gives good bounds for any $f:\{0,1\}^n\rightarrow \CN$ whose 
variance is not very large. On the other hand, Theorem \ref{main} is specific to $f(x) = e^{i\theta |x|}$ (it can also 
be used to obtain good bounds for symmetric functions  (i.e., $f(x)$ depends on the weight $|x|$ of $x$) with small $L_\infty$-norms).

\section{Comparison with random codes}\label{discr}
In this section, we compare  the bound in Corollary \ref{maincor}  on the average $L_1$-distance $E_{u\sim U_n} \| \overline{\mu_{Q+u}} - \bind_n \|_1$, 
over the random choice of coset $Q+u$,  
to the average $L_1$-distance $E_Q \| \overline{\mu_{Q+u}} - \bind_n \|_1$, when $u\in \F_2^n$ is fixed and the code 
$Q$ is chosen at random.    We focus on small codes, and namely codes of   polynomial-size.

Let $Q\subset \F_2^n$ be a random $\F_2$-linear code of size $N$ (where $N$ is a power of $2$).  
Then $|E_Q \overline{\mu_{Q}}(w) - \bind_n(w)|   \leq 1/N$ 
for $w=0,\ldots,n$ (see \cite{MS77}, page 287).  
More generally,  it is not hard to establish the following estimates. 
\begin{lemma}\label{avgl1}
Let $N$ be function of $n$ such that $N$ is a power of $2$, $N=w(1)$,  and $N = o(2^n)$. Fix any $u \in \F_2^n$. 
Let $\Gamma = E_Q\|\overline{\mu_{Q+u}}-\bind_n\|_2^2$, where $Q\subset \F_2^n$ is an $\F_2$-linear code of size $N$  chosen uniformly at random.  Then,  
$    \Gamma = \frac{1\pm o(1)}{N}$
and $\Gamma \leq E_Q\| \overline{\mu_{Q+u}}-\bind_n \|_1 \leq \sqrt{(n+1) \Gamma}.$
\end{lemma}
The proof is in Appendix \ref{appA}. 
Assume that code is of polynomial-size, i.e.,  $N = n^c$, where $c = \Theta(1)$. 
To compare with random codes, we need following simple variation of the Gilbert-Varshamov bound.
\begin{lemma} \label{GVvar}
Let $c >1$ be a constant such that $n^c$ is a power of $2$.  Then, for $n$ large enough, 
almost all $\F_2$-linear codes $Q\subset \F_2^n$ of size $n^c$ have dual bilateral minimum distance at least $d=\ceil{c}-1$. 
\end{lemma} 
The proof of Lemma \ref{GVvar} is below.  The following bound follows from Lemma \ref{GVvar} and Corollary \ref{maincor}. 
\begin{corollary}\label{nccx}  
Let $c \geq 8$ be a constant such $\ceil{c}$ is even and  $N = n^c$ is a power of $2$.  Then, for each $\e>0$,  for $n$ large enough and for 
almost all $\F_2$-linear codes $Q\subset \F_2^n$ of size $N$, we have  
 $E_{u\sim U_n} \| m_{Q+u} - \bind_n \|_1  =O(\frac{n^{1.5+\e}}{N^{1/4}})$. 
\end{corollary}
{\bf Proof:} By Lemma \ref{GVvar}, 
 for $n$ large enough,  almost all codes $Q\subset \F_2^n$ of size $n^c$ have dual bilateral minimum distance at least $d=\ceil{c}-1$.
Since $\ceil{c}\geq 8$ is even, 
 $d=\ceil{c}-1=2t+1$, for some integer $t\geq 3$.  It follows from Corollary \ref{maincor} that 
 $E_{u\sim U_n} \| m_{Q+u} - \bind_n \|_1  =O( \frac{(\ln{n})^t}{n^{t/2-1}} ) = O(\frac{n^{1.5+\e}}{N^{1/4}})$ because 
$t/2-1 = \ceil{c}/4-1.5 \geq c/4 -1.5$.
\finito

It is appropriate to   compare  the upper bound $E_{u\sim U_n} \| m_{Q+u} - \bind_n \|_1  =O(\frac{n^{1.5+\e}}{N^{1/4}})$ to 
the lower bound $E_Q\| \overline{\mu_{Q+u}}-\bind_n \|_1 \geq \Gamma = \Theta(\frac{1}{N})$ and the 
upper bound $E_Q\| \overline{\mu_{Q+u}}-\bind_n \|_1 \leq \sqrt{(n+1) \Gamma }=  \Theta(\frac{n^{1/2}}{N^{1/2}})$ of random codes. 
 The bounds differ by the  $O(n^{1.5+\e})$ factor and the exponent of $N$. 
The exponent of $N$ in Corollary \ref{nccx} comes from 
the exponent $\frac{t}{2}$  
in  Corollary \ref{maincor}.  
It is not clear what is the optimal exponent; we leave the question open.

~\\
\noindent 
{\bf Proof of Lemma \ref{GVvar}:} Note that $d\geq 1$ since $c>1$. 
Choose the generator matrix $G_{k\times n}$ of the dual code $Q^\bot$ uniformly at random, where $k = n - c\log_2{n}$.
The probability $p$ that there  exists a nonzero $x\in \F_2^n$ such that the weight of $xG$ is less than $d$ or larger than $n-d$ is 
at most $2\times (2^{k}-1) \times V(d-1)/2^n$, where $V(d-1)$ is the volume of the Hamming ball of radius $d-1$. 
We have  $V(d-1) \leq n^{d-1}+1\leq 2n^{d-1}$, thus $p \leq 4n^{-c}n^{d-1} = 4n^{-c+\ceil{c}-2} < n^{-1}$, for $n$ large enough. 
\finito

\section*{Acknowledgments} 
The author would like to thank the anonymous reviewers for their useful and constructive comments.

\newpage  
\appendix

\noindent 
{\Large \bf Appendix}
\section{Proof of Lemma \ref{avgl1}}\label{appA}
For convenience we repeat the statement of Lemma \ref{avgl1} here.

~\\
\noindent 
{\bf Lemma \ref{avgl1}}~{\em  
Let $N$ be function of $n$ such that $N$ is a power of $2$, $N=w(1)$,  and $N = o(2^n)$. Fix any $u \in \F_2^n$. 
Let $\Gamma = E_Q\|\overline{\mu_{Q+u}}-\bind_n\|_2^2$, where $Q\subset \F_2^n$ is an $\F_2$-linear code of size $N$  chosen uniformly at random.  Then,  
$    \Gamma = \frac{1\pm o(1)}{N}$
and $\Gamma \leq E_Q\| \overline{\mu_{Q+u}}-\bind_n \|_1 \leq \sqrt{(n+1) \Gamma}.$
}

~\\   
{\bf Proof:} 
The bound  $E_Q\| \overline{\mu_{Q+u}}-\bind_n \|_1 \leq \sqrt{(n+1) \Gamma}$ follows from 
Jensen's inequality applied to $g(Q,w)=\overline{\mu_{Q+u}}(w)-\bind_n(w)$ 
($(E_{w,Q} |g(Q,w)|)^2 \leq E_{w,Q} |g(Q,w)|^2$)
, and the bound  $E_Q\| \overline{\mu_{Q+u}}-\bind_n \|_1  \geq \Gamma$ follows from 
the fact that for each $Q$, 
$\|\overline{\mu_{Q+u}}(w)-\bind_n(w)\|_1 \geq \|\overline{\mu_{Q+u}}(w)-\bind_n(w)\|_2^2$ because 
$|\overline{\mu_{Q+u}}(w)-\bind_n(w)|\leq 1$ for each $w\in [0: n]$.

To establish the estimate $\Gamma = \frac{1\pm o(1)}{N}$, for each $w\in [0:n]$, define $f_w:\{0,1\}^n \rightarrow \R$ by $f_w(y) = I_w(y+u) - \bind_n(w)$, where 
$I_w:\{0,1\}^n \rightarrow \{0,1\}$ is the indicator function given by $I_w(x)=1$ iff $|x|=w$.  
Thus, $\overline{\mu_{Q+u}}(w)-\bind_n(w) = E_{y\sim \mu_{Q}} f_w(y)$ 
and $\Gamma= \sum_w \Gamma_w$, where  $\Gamma_w = E_Q (E_{y\sim \mu_{Q}} f_w(y))^2$.

Fix any $w\in [0:n]$ and note that  $E_{U_n} f_w  =0$.
We have $$(E_{y\sim \mu_{Q}} f_w(y))^2
 = E_{y,y'\sim \mu_{Q}} f_w(y)f_w(y') = \frac{1}{N^2}\sum_{y,y'\in Q} f_w(y) f_w(y').$$ 
Each nonzero $y\in \{0,1\}^n$ belongs to $Q$ with probability $p_N \defeq \frac{N-1}{2^n-1}$.
The $y =0 $ case is special as $0$ must be in  the code.   Moreover, for $N \geq 4$,
for any distinct $y,y'\neq 0$, the events $\{y\in Q\}$  and $\{y'\in Q\}$  are independent.  Therefore, 
\begin{eqnarray}
\Gamma_w &=&   E_Q (E_{y\sim \mu_{Q}} f_w(y))^2 \nonumber \\
&=&
 \frac{1}{N^2}\left(\sum_{y,y'\neq 0 : y\neq y'} p_N^2f_w(y) f_w(y')+\sum_{y\neq 0 } p_N f_w(y)^2 +2\sum_{y\neq 0 } p_N f_w(y)f_w(0) + f_w(0)^2\right).\label{messeq}
\end{eqnarray}
Since $E_{U_n}f_w =0$, we  have $\sum_{y\neq 0} f_w(y)  = - f_w(0)$ . Moreover, 
$\frac{1}{2^{2n}}\sum_{y,y'} f_w(y) f_w(y') = \left(E_{U_n}f_w\right)^2=0$, thus   
\[
    \sum_{y,y'\neq 0 : y\neq y'} f_w(y) f_w(y') = - \sum_{y\neq 0 }  f_w(y)^2 -2\sum_{y\neq 0 } f_w(y)f_w(0) - f_w(0)^2. 
\]
Replacing  in (\ref{messeq}), we obtain 
\begin{eqnarray*}
    \Gamma_w &=&  \left(\frac{p_N}{N^2}-\left(\frac{p_N}{N}\right)^2  \right)\sum_{y\neq 0 }  f_w(y)^2 +2\left(\frac{p_N}{N^2}-\left(\frac{p_N}{N}\right)^2 \right)\sum_{y\neq 0 } f_w(y)f_w(0) + \left(\frac{1}{N^2}-\left(\frac{p_N}{N}\right)^2 \right)f_w(0)^2 \\
&=&  \left(\frac{p_N}{N^2}-\left(\frac{p_N}{N}\right)^2  \right)
\sum_{y}  f_w(y)^2 -2\left(\frac{p_N}{N^2}-\left(\frac{p_N}{N}\right)^2 \right)f_w(0)^2 + \left(\frac{1}{N^2}-\frac{p_N}{N^2}
 \right)f_w(0)^2 \\
    &=&  \frac{a}{N}
E_{U_n}  f_w^2 + \frac{b}{N^2}f_w(0)^2,
\end{eqnarray*}
where 
$a = \frac{2^n}{N}(p_N-p_N^2)=\left(1-\frac{1}{N}\right)\frac{1-p_N}{1-2^{-n}}$, and  $b =  1+2p_N^2   -3p_N$.  
Note that  since  $N = w(1)$ and $N = o(2^n)$, and hence $p_N = o(1)$, 
we have $a = 1 -o(1)$ and $b = 1 - o(1)$.  Now, $$E_{U_n}  f_w^2 = E_{y\sim U_n} I_w(y+u)^2 - \bind_n(w)^2= \bind_n(w) - \bind_n(w)^2$$ and 
$f_w(0)^2 = (I_w(u)-\bind_n(w))^2$. 
It follows that $\sum_w E_{U_n}  f_w^2 =  1 - \sum_w \bind_n(w)^2$ and 
$$\sum_w f_w(0)^2 = (1-\bind_n(|u|))^2 + \sum_{w\neq |u|} \bind_n(w)^2 = 1 - 2 \bind_n(|u|) + \sum_w \bind_n(w)^2.$$ 
Using the identity, $\sum_w \binom{n}{w}^2 = \binom{2n}{n}$, we get 
$\sum_w \bind_n(w)^2 = \bind_{2n}(n)$.  
Therefore, 
\begin{eqnarray*}
    \Gamma     &=&  \frac{a}{N}
\left(1 - \bind_{2n}(n)\right)
 + 
\frac{b}{N^2}
\left(1 - 2 \bind_n(|u|) +  \bind_{2n}(n)\right)
 \\
&=& \frac{a}{N}
\left(1 - O(n^{-1/2})\right)
 + 
\frac{b}{N^2}
\left(1 \pm O(n^{-1/2})\right)
\\
&=& \frac{1\pm o(1)}{N}.
\end{eqnarray*}
\finito

\end{document}